\documentclass[letter,longauth]{aa}
\usepackage{graphicx}
\usepackage{txfonts}
\usepackage[colorlinks=true, allcolors=blue]{hyperref}
\definecolor{Red}{rgb}{0.9,0,0}
\definecolor{Blue}{rgb}{0,0,0.9}
\definecolor{Green}{rgb}{0,0.5,0}
\definecolor{Black}{rgb}{0,0,0}

\usepackage{natbib}
\bibpunct{(}{)}{;}{a}{}{,} 

\newcommand{\Autoref}[1]{%
  \begingroup%
  \def\chapterautorefname{Chapter}%
  \def\sectionautorefname{Section}%
  \def\subsectionautorefname{Subsection}%
  \def\subsubsectionautorefname{Subsubsection}%
  \def\paragraphautorefname{Paragraph}%
  \def\tableautorefname{Table}%
  \def\equationautorefname{Equation}%
  \autoref{#1}%
  \endgroup%
}

\begin{document}

\title{A stellar occultation by the transneptunian object (50000) Quaoar observed by CHEOPS\thanks{This article uses data from CHEOPS programme CH\_PR100021.}
}
\titlerunning{Stellar occultation observed by CHEOPS}

\authorrunning{B. Morgado, G. Bruno, A. R. Gomes-J\'unior etal.,}

\author{
B. E. Morgado\inst{\ref{inst:a},\ref{inst:b},\ref{inst:c}} $^{\href{https://orcid.org/0000-0003-0088-1808}{\includegraphics[scale=0.005]{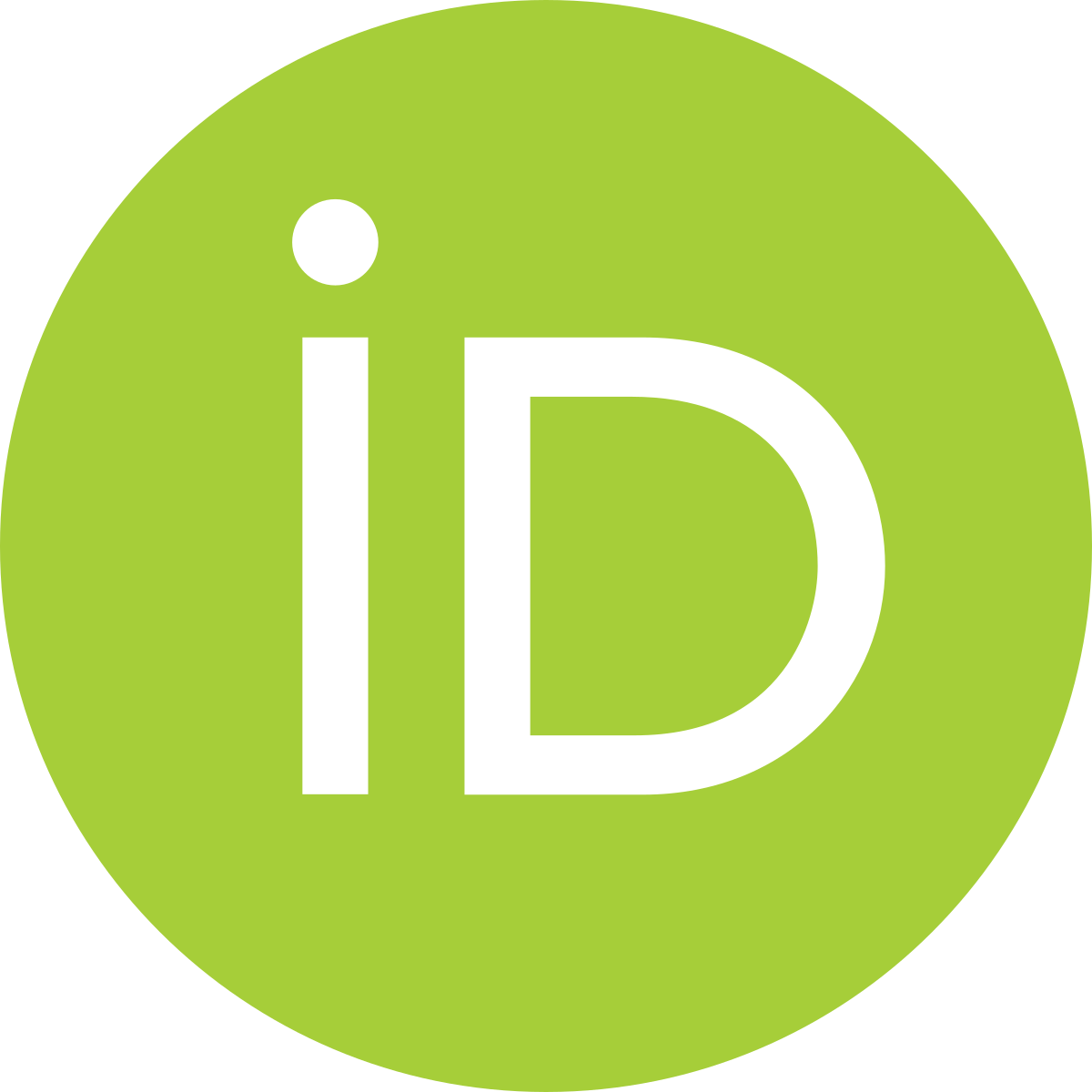}}}$ 
\and G. Bruno\inst{\ref{inst:1}} $^{\href{https://orcid.org/0000-0002-3288-0802}{\includegraphics[scale=0.005]{Figs/orcid.png}}}$
\and A. R. Gomes-Júnior\inst{\ref{inst:z},\ref{inst:d},\ref{inst:c}}  $^{\href{https://orcid.org/0000-0002-3362-2127}{\includegraphics[scale=0.005]{Figs/orcid.png}}}$ 
\and I. Pagano\inst{\ref{inst:1}} $^{\href{https://orcid.org/0000-0001-9573-4928}{\includegraphics[scale=0.005]{Figs/orcid.png}}}$
\and B. Sicardy\inst{\ref{inst:e}}  $^{\href{https://orcid.org/0000-0003-1995-0842}{\includegraphics[scale=0.005]{Figs/orcid.png}}}$
\and A. Fortier\inst{\ref{inst:6},\ref{inst:16}} $^{\href{https://orcid.org/0000-0001-8450-3374}{\includegraphics[scale=0.005]{Figs/orcid.png}}}$
\and J. Desmars\inst{\ref{inst:f},\ref{inst:g}}  $^{\href{https://orcid.org/0000-0002-2193-8204}{\includegraphics[scale=0.005]{Figs/orcid.png}}}$
\and P. F. L. Maxted\inst{\ref{inst:35}} $^{\href{https://orcid.org/0000-0003-3794-1317}{\includegraphics[scale=0.005]{Figs/orcid.png}}}$
\and F. Braga-Ribas\inst{\ref{inst:h},\ref{inst:b},\ref{inst:c}}  $^{\href{https://orcid.org/0000-0003-2311-2438}{\includegraphics[scale=0.005]{Figs/orcid.png}}}$
\and D. Queloz\inst{\ref{inst:39},\ref{inst:40}} $^{\href{https://orcid.org/0000-0002-3012-0316}{\includegraphics[scale=0.005]{Figs/orcid.png}}}$
\and S. G. Sousa\inst{\ref{inst:13}} $^{\href{https://orcid.org/0000-0001-9047-2965}{\includegraphics[scale=0.005]{Figs/orcid.png}}}$
\and J. L. Ortiz\inst{\ref{inst:l}} $^{\href{https://orcid.org/0000-0002-8690-2413}{\includegraphics[scale=0.005]{Figs/orcid.png}}}$
\and A. Brandeker\inst{\ref{inst:2}} $^{\href{https://orcid.org/0000-0002-7201-7536}{\includegraphics[scale=0.005]{Figs/orcid.png}}}$
\and A. Collier Cameron\inst{\ref{inst:3}} $^{\href{https://orcid.org/0000-0002-8863-7828}{\includegraphics[scale=0.005]{Figs/orcid.png}}}$
\and C. L. Pereira\inst{\ref{inst:b},\ref{inst:c}}  $^{\href{https://orcid.org/0000-0003-1000-8113}{\includegraphics[scale=0.005]{Figs/orcid.png}}}$
\and H. G. Florén\inst{\ref{inst:2}}
\and N. Hara\inst{\ref{inst:4}} $^{\href{https://orcid.org/0000-0001-9232-3314}{\includegraphics[scale=0.005]{Figs/orcid.png}}}$
\and D. Souami\inst{\ref{inst:47},\ref{inst:e},\ref{inst:i}}  $^{\href{https://orcid.org/0000-0003-4058-0815}{\includegraphics[scale=0.005]{Figs/orcid.png}}}$
\and K. G. Isaak\inst{\ref{inst:5}} $^{\href{https://orcid.org/0000-0001-8585-1717}{\includegraphics[scale=0.005]{Figs/orcid.png}}}$
\and G. Olofsson\inst{\ref{inst:2}} $^{\href{https://orcid.org/0000-0003-3747-7120}{\includegraphics[scale=0.005]{Figs/orcid.png}}}$
\and P. Santos-Sanz\inst{\ref{inst:l}} $^{\href{https://orcid.org/0000-0002-1123-983X}{\includegraphics[scale=0.005]{Figs/orcid.png}}}$
\and T. G. Wilson\inst{\ref{inst:3}} $^{\href{https://orcid.org/0000-0001-8749-1962}{\includegraphics[scale=0.005]{Figs/orcid.png}}}$
\and J. Broughton\inst{\ref{inst:j},\ref{inst:k}} 
\and Y. Alibert\inst{\ref{inst:6}} $^{\href{https://orcid.org/0000-0002-4644-8818}{\includegraphics[scale=0.005]{Figs/orcid.png}}}$
\and R. Alonso\inst{\ref{inst:7}, \ref{inst:8}} $^{\href{https://orcid.org/0000-0001-8462-8126}{\includegraphics[scale=0.005]{Figs/orcid.png}}}$
\and G. Anglada\inst{\ref{inst:9},\ref{inst:10}} $^{\href{https://orcid.org/0000-0002-3645-5977}{\includegraphics[scale=0.005]{Figs/orcid.png}}}$
\and T. Bárczy\inst{\ref{inst:11}} $^{\href{https://orcid.org/0000-0002-7822-4413}{\includegraphics[scale=0.005]{Figs/orcid.png}}}$
\and D. Barrado\inst{\ref{inst:12}} $^{\href{https://orcid.org/0000-0002-5971-9242}{\includegraphics[scale=0.005]{Figs/orcid.png}}}$
\and S. C. C. Barros\inst{\ref{inst:13},\ref{inst:14}} $^{\href{https://orcid.org/0000-0003-2434-3625}{\includegraphics[scale=0.005]{Figs/orcid.png}}}$
\and W. Baumjohann\inst{\ref{inst:15}} $^{\href{https://orcid.org/0000-0001-6271-0110}{\includegraphics[scale=0.005]{Figs/orcid.png}}}$
\and M. Beck\inst{\ref{inst:4}} $^{\href{https://orcid.org/0000-0003-3926-0275}{\includegraphics[scale=0.005]{Figs/orcid.png}}}$
\and T. Beck\inst{\ref{inst:6}}
\and W. Benz\inst{\ref{inst:6}, \ref{inst:16}} $^{\href{https://orcid.org/0000-0001-7896-6479}{\includegraphics[scale=0.005]{Figs/orcid.png}}}$
\and N. Billot\inst{\ref{inst:4}} $^{\href{https://orcid.org/0000-0003-3429-3836}{\includegraphics[scale=0.005]{Figs/orcid.png}}}$
\and X. Bonfils\inst{\ref{inst:17}} $^{\href{https://orcid.org/0000-0001-9003-8894}{\includegraphics[scale=0.005]{Figs/orcid.png}}}$
\and C. Broeg\inst{\ref{inst:6}, \ref{inst:16}} $^{\href{https://orcid.org/0000-0001-5132-2614}{\includegraphics[scale=0.005]{Figs/orcid.png}}}$
\and J. Cabrera\inst{\ref{inst:19}}
\and S. Charnoz\inst{\ref{inst:20}} $^{\href{https://orcid.org/0000-0002-7442-491X}{\includegraphics[scale=0.005]{Figs/orcid.png}}}$
\and S. Csizmadia\inst{\ref{inst:19}} $^{\href{https://orcid.org/0000-0001-6803-9698}{\includegraphics[scale=0.005]{Figs/orcid.png}}}$
\and M. B. Davies\inst{\ref{inst:21}} $^{\href{https://orcid.org/0000-0001-6080-1190}{\includegraphics[scale=0.005]{Figs/orcid.png}}}$
\and M. Deleuil\inst{\ref{inst:22}} $^{\href{https://orcid.org/0000-0001-6036-0225}{\includegraphics[scale=0.005]{Figs/orcid.png}}}$
\and L. Delrez\inst{\ref{inst:23},\ref{inst:24}} $^{\href{https://orcid.org/0000-0001-6108-4808}{\includegraphics[scale=0.005]{Figs/orcid.png}}}$
\and O. D. S. Demangeon\inst{\ref{inst:13},\ref{inst:14}} $^{\href{https://orcid.org/0000-0001-7918-0355}{\includegraphics[scale=0.005]{Figs/orcid.png}}}$
\and B. O. Demory\inst{\ref{inst:16}} $^{\href{https://orcid.org/0000-0002-9355-5165}{\includegraphics[scale=0.005]{Figs/orcid.png}}}$
\and D. Ehrenreich\inst{\ref{inst:4}} $^{\href{https://orcid.org/0000-0001-9704-5405}{\includegraphics[scale=0.005]{Figs/orcid.png}}}$
\and A. Erikson\inst{\ref{inst:19}}
\and L. Fossati\inst{\ref{inst:15}} $^{\href{https://orcid.org/0000-0003-4426-9530}{\includegraphics[scale=0.005]{Figs/orcid.png}}}$
\and M. Fridlund\inst{\ref{inst:25},\ref{inst:26}} $^{\href{https://orcid.org/0000-0002-0855-8426}{\includegraphics[scale=0.005]{Figs/orcid.png}}}$
\and D. Gandolfi\inst{\ref{inst:27}} $^{\href{https://orcid.org/0000-0001-8627-9628}{\includegraphics[scale=0.005]{Figs/orcid.png}}}$
\and M. Gillon\inst{\ref{inst:23}} $^{\href{https://orcid.org/0000-0003-1462-7739}{\includegraphics[scale=0.005]{Figs/orcid.png}}}$
\and M. Güdel\inst{\ref{inst:28}}
\and K. Heng\inst{\ref{inst:16},\ref{inst:29}} $^{\href{https://orcid.org/0000-0003-1907-5910}{\includegraphics[scale=0.005]{Figs/orcid.png}}}$
\and S. Hoyer\inst{\ref{inst:22}} $^{\href{https://orcid.org/0000-0003-3477-2466}{\includegraphics[scale=0.005]{Figs/orcid.png}}}$
\and L. L. Kiss\inst{\ref{inst:30},\ref{inst:31}}
\and J. Laskar\inst{\ref{inst:32}} $^{\href{https://orcid.org/0000-0003-2634-789X}{\includegraphics[scale=0.005]{Figs/orcid.png}}}$
\and A. Lecavelier des Etangs\inst{\ref{inst:33}} $^{\href{https://orcid.org/0000-0002-5637-5253}{\includegraphics[scale=0.005]{Figs/orcid.png}}}$
\and M. Lendl\inst{\ref{inst:4}} $^{\href{https://orcid.org/0000-0001-9699-1459}{\includegraphics[scale=0.005]{Figs/orcid.png}}}$
\and C. Lovis\inst{\ref{inst:4}} $^{\href{https://orcid.org/0000-0001-7120-5837}{\includegraphics[scale=0.005]{Figs/orcid.png}}}$
\and D. Magrin\inst{\ref{inst:34}} $^{\href{https://orcid.org/0000-0003-0312-313X}{\includegraphics[scale=0.005]{Figs/orcid.png}}}$
\and L. Marafatto\inst{\ref{inst:34}}
\and V. Nascimbeni\inst{\ref{inst:34}} $^{\href{https://orcid.org/0000-0001-9770-1214}{\includegraphics[scale=0.005]{Figs/orcid.png}}}$
\and R. Ottensamer\inst{\ref{inst:36}} $^{\href{https://orcid.org/0000-0001-5684-5836}{\includegraphics[scale=0.005]{Figs/orcid.png}}}$
\and E. Pallé\inst{\ref{inst:7}} $^{\href{https://orcid.org/0000-0003-0987-1593}{\includegraphics[scale=0.005]{Figs/orcid.png}}}$
\and G. Peter\inst{\ref{inst:37}} $^{\href{https://orcid.org/0000-0001-6101-2513}{\includegraphics[scale=0.005]{Figs/orcid.png}}}$
\and D. Piazza\inst{\ref{inst:6}}
\and G. Piotto\inst{\ref{inst:34},\ref{inst:38}} $^{\href{https://orcid.org/0000-0002-9937-6387}{\includegraphics[scale=0.005]{Figs/orcid.png}}}$
\and D. Pollacco\inst{\ref{inst:29}}
\and R. Ragazzoni\inst{\ref{inst:34},\ref{inst:38}} $^{\href{https://orcid.org/0000-0002-7697-5555}{\includegraphics[scale=0.005]{Figs/orcid.png}}}$
\and N. Rando\inst{\ref{inst:41}}
\and F. Ratti\inst{\ref{inst:41}}
\and H. Rauer\inst{\ref{inst:19},\ref{inst:42},\ref{inst:43}} $^{\href{https://orcid.org/0000-0002-6510-1828}{\includegraphics[scale=0.005]{Figs/orcid.png}}}$
\and C. Reimers\inst{\ref{inst:36}} $^{\href{https://orcid.org/0000-0002-2334-1620}{\includegraphics[scale=0.005]{Figs/orcid.png}}}$
\and I. Ribas\inst{\ref{inst:9},\ref{inst:10}} $^{\href{https://orcid.org/0000-0002-6689-0312}{\includegraphics[scale=0.005]{Figs/orcid.png}}}$
\and N. C. Santos\inst{\ref{inst:13},\ref{inst:14}} $^{\href{https://orcid.org/0000-0003-4422-2919}{\includegraphics[scale=0.005]{Figs/orcid.png}}}$
\and G. Scandariato\inst{\ref{inst:1}} $^{\href{https://orcid.org/0000-0003-2029-0626}{\includegraphics[scale=0.005]{Figs/orcid.png}}}$
\and D. Ségransan\inst{\ref{inst:4}} $^{\href{https://orcid.org/0000-0003-2355-8034}{\includegraphics[scale=0.005]{Figs/orcid.png}}}$
\and A. E. Simon\inst{\ref{inst:6}} $^{\href{https://orcid.org/0000-0001-9773-2600}{\includegraphics[scale=0.005]{Figs/orcid.png}}}$
\and A. M. S. Smith\inst{\ref{inst:19}} $^{\href{https://orcid.org/0000-0002-2386-4341}{\includegraphics[scale=0.005]{Figs/orcid.png}}}$
\and M. Steller\inst{\ref{inst:15}} $^{\href{https://orcid.org/0000-0003-2459-6155}{\includegraphics[scale=0.005]{Figs/orcid.png}}}$
\and G. M. Szabó\inst{\ref{inst:44},\ref{inst:45}}
\and N. Thomas\inst{\ref{inst:6}}
\and S. Udry\inst{\ref{inst:4}} $^{\href{https://orcid.org/0000-0001-7576-6236}{\includegraphics[scale=0.005]{Figs/orcid.png}}}$
\and V. Van Grootel\inst{\ref{inst:24}} $^{\href{https://orcid.org/0000-0003-2144-4316}{\includegraphics[scale=0.005]{Figs/orcid.png}}}$
\and N. A. Walton\inst{\ref{inst:46}} $^{\href{https://orcid.org/0000-0003-3983-8778}{\includegraphics[scale=0.005]{Figs/orcid.png}}}$
\and K. Westerdorff\inst{\ref{inst:37}}
}

\institute{
\label{inst:a} Universidade Federal do Rio de Janeiro - Observatório do Valongo, Ladeira Pedro Antônio 43, CEP 20.080-090 Rio de Janeiro - RJ, Brazil\\
\email{bmorgado@ov.ufrj.br} 
\and \label{inst:b} Observatório Nacional/MCTI, R. General José Cristino 77, CEP 20921-400 Rio de Janeiro - RJ, Brazil 
\and \label{inst:c} Laboratório Interinstitucional de e-Astronomia - LIneA, Rua Gal. José Cristino 77, Rio de Janeiro, RJ 20921-400, Brazil 
\and \label{inst:1} INAF, Osservatorio Astrofisico di Catania, Via S. Sofia 78, 95123 Catania, Italy
\and \label{inst:z} Institute of Physics, Federal University of Uberlândia, Uberlândia-MG, Brazil 
\and \label{inst:d} UNESP - São Paulo State University, Grupo de Dinâmica Orbital e Planetologia, CEP 12516-410, Guaratinguetá, SP, Brazil 
\and \label{inst:e} LESIA, Observatoire de Paris, Université PSL, Sorbonne Université, Université de Paris, CNRS, 92190 Meudon, France 
\and \label{inst:6} Physikalisches Institut, University of Bern, Sidlerstrasse 5, 3012 Bern, Switzerland 
\and \label{inst:16} Center for Space and Habitability, University of Bern, Gesellschaftsstrasse 6, 3012 Bern, Switzerland
\and \label{inst:f} Institut Polytechnique des Sciences Avancées IPSA, 63 boulevard de Brandebourg, 94200 Ivry-sur-Seine, France 
\and \label{inst:g} Institut de Mécanique Céleste et de Calcul des Éphémérides, IMCCE, Observatoire de Paris, PSL Research University, CNRS,Sorbonne Universités, UPMC Univ Paris 06, Univ. Lille, 77, Av. Denfert-Rochereau, 75014 Paris, France 
\and \label{inst:35} Astrophysics Group, Keele University, Staffordshire, ST5 5BG, United Kingdom
\and \label{inst:h} Federal University of Technology - Paraná (UTFPR / DAFIS), Rua Sete de Setembro, 3165, CEP 80230-901, Curitiba, PR, Brazil 
\and \label{inst:39} ETH Zurich, Department of Physics, Wolfgang-Pauli-Strasse 2, CH-8093 Zurich, Switzerland
\and \label{inst:40} Cavendish Laboratory, JJ Thomson Avenue, Cambridge CB3 0HE, UK
\and \label{inst:13} Instituto de Astrofisica e Ciencias do Espaco, Universidade do Porto, CAUP, Rua das Estrelas, 4150-762 Porto, Portugal
\and \label{inst:l} Instituto de Astrofísica de Andalucía, IAA-CSIC, Glorieta de la Astronomía s/n, 18008 Granada, Spain 
\and \label{inst:2} Department of Astronomy, Stockholm University, AlbaNova University Center, 10691 Stockholm, Sweden \and
\label{inst:3} Centre for Exoplanet Science, SUPA School of Physics and Astronomy, University of St Andrews, North Haugh, St Andrews KY16 9SS, UK 
\and \label{inst:4} Observatoire Astronomique de l'Université de Genève, Chemin Pegasi 51, Versoix, Switzerland 
\and \label{inst:47} Observatoire de la Côte d'Azur, Laboratoire Lagrange UMR7293 CNRS, Nice, France
\and  \label{inst:i} naXys, University of Namur, 8 Rempart de la Vierge, Namur, B-5000, Belgium 
\and \label{inst:5} Science and Operations Department - Science Division (SCI-SC), Directorate of Science, European Space Agency (ESA), European Space Research and Technology Centre (ESTEC), Keplerlaan 1, 2201-AZ Noordwijk, The Netherlands 
\and \label{inst:j} Reedy Creek Observatory, Gold Coast, Queensland, Australia 
\and \label{inst:k} Trans-Tasman Occultation Alliance (TTOA), Wellington, PO Box 3181, New Zealand 
\and \label{inst:7} Instituto de Astrofisica de Canarias, 38200 La Laguna, Tenerife, Spain
\and \label{inst:8} Departamento de Astrofisica, Universidad de La Laguna, 38206 La Laguna, Tenerife, Spain
\and \label{inst:9} Institut de Ciencies de l'Espai (ICE, CSIC), Campus UAB, Can Magrans s/n, 08193 Bellaterra, Spain
\and \label{inst:10} Institut d'Estudis Espacials de Catalunya (IEEC), 08034 Barcelona, Spain
\and \label{inst:11} Admatis, 5. Kandó Kálmán Street, 3534 Miskolc, Hungary
\and \label{inst:12} Depto. de Astrofisica, Centro de Astrobiologia (CSIC-INTA), ESAC campus, 28692 Villanueva de la Cañada (Madrid), Spain
\and \label{inst:14} Departamento de Fisica e Astronomia, Faculdade de Ciencias, Universidade do Porto, Rua do Campo Alegre, 4169-007 Porto, Portugal
\and \label{inst:15} Space Research Institute, Austrian Academy of Sciences, Schmiedlstrasse 6, A-8042 Graz, Austria
\and \label{inst:17} Université Grenoble Alpes, CNRS, IPAG, 38000 Grenoble, France
\and \label{inst:19} Institute of Planetary Research, German Aerospace Center (DLR), Rutherfordstrasse 2, 12489 Berlin, Germany
\and \label{inst:20} Université de Paris, Institut de physique du globe de Paris, CNRS, F-75005 Paris, France
\and \label{inst:21} Centre for Mathematical Sciences, Lund University, Box 118, 221 00 Lund, Sweden
\and \label{inst:22} Aix Marseille Univ, CNRS, CNES, LAM, 38 rue Frédéric Joliot-Curie, 13388 Marseille, France
\and \label{inst:23} Astrobiology Research Unit, Université de Liège, Allée du 6 Août 19C, B-4000 Liège, Belgium
\and \label{inst:24} Space sciences, Technologies and Astrophysics Research (STAR) Institute, Université de Liège, Allée du 6 Août 19C, 4000 Liège, Belgium
\and \label{inst:25} Leiden Observatory, University of Leiden, PO Box 9513, 2300 RA Leiden, The Netherlands
\and \label{inst:26} Department of Space, Earth and Environment, Chalmers University of Technology, Onsala Space Observatory, 439 92 Onsala, Sweden
\and \label{inst:27} Dipartimento di Fisica, Universita degli Studi di Torino, via Pietro Giuria 1, I-10125, Torino, Italy
\and \label{inst:28} University of Vienna, Department of Astrophysics, Türkenschanzstrasse 17, 1180 Vienna, Austria
\and \label{inst:29} Department of Physics, University of Warwick, Gibbet Hill Road, Coventry CV4 7AL, United Kingdom
\and \label{inst:30} Konkoly Observatory, Research Centre for Astronomy and Earth Sciences, 1121 Budapest, Konkoly Thege Miklós út 15-17, Hungary
\and \label{inst:31} ELTE E\"otv\"os Lor\'and University, Institute of Physics, P\'azm\'any P\'eter s\'et\'any 1/A, 1117
\and \label{inst:32} IMCCE, UMR8028 CNRS, Observatoire de Paris, PSL Univ., Sorbonne Univ., 77 av. Denfert-Rochereau, 75014 Paris, France
\and \label{inst:33} Institut d'astrophysique de Paris, UMR7095 CNRS, Université Pierre \& Marie Curie, 98bis blvd. Arago, 75014 Paris, France
\and \label{inst:34} INAF, Osservatorio Astronomico di Padova, Vicolo dell'Osservatorio 5, 35122 Padova, Italy
\and \label{inst:36} Department of Astrophysics, University of Vienna, Tuerkenschanzstrasse 17, 1180 Vienna, Austria
\and \label{inst:37} Institute of Optical Sensor Systems, German Aerospace Center (DLR), Rutherfordstrasse 2, 12489 Berlin, Germany
\and \label{inst:38} Dipartimento di Fisica e Astronomia "Galileo Galilei", Universita degli Studi di Padova, Vicolo dell'Osservatorio 3, 35122 Padova, Italy
\and \label{inst:41} ESTEC, European Space Agency, 2201AZ, Noordwijk, NL
\and \label{inst:42} Zentrum für Astronomie und Astrophysik, Technische Universität Berlin, Hardenbergstr. 36, D-10623 Berlin, Germany
\and \label{inst:43} Institut für Geologische Wissenschaften, Freie Universität Berlin, 12249 Berlin, Germany
\and \label{inst:44} ELTE E\"otv\"os Lor\'and University, Gothard Astrophysical Observatory, 9700 Szombathely, Szent Imre h. u. 112, Hungary
\and \label{inst:45} MTA-ELTE Exoplanet Research Group, 9700 Szombathely, Szent Imre h. u. 112, Hungary
\and \label{inst:46} Institute of Astronomy, University of Cambridge, Madingley Road, Cambridge, CB3 0HA, United Kingdom
}
   \date{Received 09/06/2022; accepted 09/08/2022}

 
  \abstract
   {Stellar occultation is a powerful technique that allows the determination of some physical parameters of the occulting object. The result depends on the photometric accuracy, the temporal resolution, and the number of chords obtained. Space telescopes can achieve high photometric accuracy as they are not affected by atmospheric scintillation.}
   {Using ESA's CHEOPS space telescope, we observed a stellar occultation by the transneptunian object (50000) Quaoar. We compare the obtained chord with previous occultations by this object and determine its astrometry with sub-milliarcsecond precision. Also, we determine upper limits to the presence of a global methane atmosphere on the occulting body.}
   {We predicted and observed a stellar occultation by Quaoar using the CHEOPS space telescope. We measured the occultation light curve from this dataset and determined the dis- and reappearance of the star behind the occulting body. Furthermore, a ground-based telescope in Australia was used to constrain Quaoar's limb. Combined with results from previous works, these measurements allowed us to obtain a precise position of Quaoar at the occultation time.}
   {We present the results obtained from the first stellar occultation by a transneptunian object (TNO) using a space telescope orbiting Earth; it was the occultation by Quaoar observed on 2020 June 11. We used the CHEOPS light curve to obtain a surface pressure upper limit of 85 nbar for the detection of a global methane atmosphere. Also, combining this observation with a ground-based observation, we fitted Quaoar's limb to determine its astrometric position with an uncertainty below 1.0 mas.}
   {This observation is the first of its kind, and it shall be considered as a proof of concept of stellar occultation observations of transneptunian objects with space telescopes orbiting Earth. Moreover, it shows significant prospects for the James Webb Space Telescope.}

   \maketitle 
%

\section{Introduction}\label{Sec:intro}

Stellar occultations happen when a body passes in front of a star as viewed by an observer. The detection of these events allow the determination of 2D apparent sizes and shapes with kilometre uncertainties \citep{Sicardy_2011}. Also, with these events, we can probe the vicinity of the occulting object in search for material, such as rings \citep{Braga-Ribas_2014, Ortiz_2017}, and even detect, characterise, or determine limits to atmospheres \citep{Marques_Oliveira_2022, Meza_2019, Arimatsu_2019, Ortiz_2012}. From an astrometric point of view, these events can provide highly accurate positions of the occulting object, with uncertainties of the order of a few milliarcseconds \citep[mas, ][]{Rommel_2020}.

This Letter details the analysis of the stellar occultation by the large transneptunian object (TNO; 50000) Quaoar on 2020 June 11. Quaoar belongs to the dynamical class of Cubewanos. It has a semi-major axis of 43.51 au, an orbital eccentricity of 0.035, and an inclination of 7.98 degrees. Quaoar was discovered in 2002 by \cite{Brown_2004}. These authors estimated Quaoar’s diameter to be about 1260 $\pm$ 190 km based on Hubble Space Telescope (HST) images. From previous stellar occultations, \cite{Braga-Ribas_2013} obtained an area equivalent diameter of 1110 $\pm$ 5.0 km under the assumption that Quaoar is a Maclaurin spheroid. Also, thermal models based on Herschel data obtained an area equivalent diameter of 1074 $\pm$ 38 km \citep{Fornasier_2013}. 

In 2007, a small moon orbiting Quaoar was discovered \citep{Brown_2007}, and was later named Weywot. This satellite orbits Quaoar with a semi-major axis of $1.45 \times 10^{4}$ km and a period of $12.047$ days. From its orbital motion, the mass of Quaoar was estimated as $1.65 \times 10^{21}$ kg \citep{Fraser_2010, Fraser_2013, Vachier_2012}.

The occultation reported in the present Letter was observable from Australia and was detected by ESA's CHaracterising ExOPlanet Satellite \citep[CHEOPS, ][]{Benz_2020} space telescope\footnote{Webpage: \url{http://www.esa.int/Science_Exploration/Space_Science/Cheops}}. CHEOPS is a mission dedicated to studying exoplanets using the transit technique. This satellite was launched in December 2019, and is in a Sun-synchronous orbit, about 700 km above Earth. It has a 32 cm diameter Ritchey-Chrétien telescope (f/8) with a single, frame-transfer, back-illuminated CCD detector in its focal plane. Its focal plane is defocussed to deliver a large point spread function (PSF), where 90\% of the stellar flux spreads within a radial aperture of 16 pixels. The CCD detector has 1024 $\times$ 1024 pixels and a pixel pitch of 13 $\mu$m.  There is no filter in the optical path, resulting in a bandpass between 0.33 and 1.10 microns. This system was designed to deliver photometric measurements with high accuracy. 

For the first time, we predicted and detected a stellar occultation by a TNO as observed by a space telescope orbiting Earth. This achievement was made possible thanks to the collaborative effort of the European Research Council (ERC) \textit{Lucky Star} project\footnote{Webpage: \url{https://lesia.obspm.fr/lucky-star/index.php}} and the ESA CHEOPS mission. Through this observation we are able to probe the object's vicinity, avoiding interference from Earth's atmosphere, detect a limit of a methane atmosphere around Quaoar, and determine the astrometric position of this large TNO with an uncertainty at the sub-mas level (less than 5 km at Quaoar's distance). 

\Autoref{sec:observation} contains details about the prediction and observation of the occultation event. In \Autoref{sec:lightcurve}, we describe the analysis of the images, determination of the light curve, and fitting the immersion (disappearance) and emersion (reappearance) times. \Autoref{sec:Limb-fit} details the geometric reconstruction of the event and a determination of Quaoar's astrometric position. Finally, our discussion and final remarks can be found in \Autoref{sec:discussion}.

\section{Prediction and observation} \label{sec:observation}

The event presented here was predicted in the framework of the ERC \textit{Lucky Star} project. It uses the star's position available in the Gaia catalogue \citep{Gaia_Mission, Gaiadr1_2016, Gaiadr2_2018}. Quaoar's ephemeris was pinned down to the 5 mas accuracy level ($\sim$150 km at Quaoar's distance), using the Numerical Integration of the Motion
of an Asteroid \citep[\textsc{nima}, ][]{Desmars_2015} integrator fed by observations in the Minor Planet Center\footnote{Webpage: \url{https://www.minorplanetcenter.net/iau/mpc.html}} (MPC) database and astrometrical positions derived from previous stellar occultations \citep{Braga-Ribas_2019, Braga-Ribas_2013}.

\Autoref{table:occinfo} contains some general information about the June 2020 occultation, such as the Gaia Source ID of the occulted star, its G magnitude,  and its position at the occultation epoch considering the robust propagation as suggested by \cite{Butkevich_2014}, also considering the correct proper motion as discussed by \cite{CantatGaudin_2021}. In this table we also added information about the occulting object, such as its positions in the occultation instant and visual magnitude.

\begin{table}[h]
\begin{center}
\caption{Information about the occultation by Quaoar on 2020 June 11.}
\vspace{4pt}
\begin{tabular}{l c} 
\hline 
\hline 
\textbf{Parameter} & \textbf{Value} \\ 
\hline
\hline 
GAIA EDR3 Source & 4146202445050212352 \\
Star's Mag G & 12.667 $\pm$ 0.003 \\
Star's RA & \phantom{-}18 15 03.055795  $\pm$ 0.09 mas  \\
Star's Dec & -15 15 04.94607\phantom{0}  $\pm$ 0.06 mas \\

Star's diameter & 0.03 mas / 0.97 km \\

Occ. date and time & 2020-06-11 16:27:25.5 UTC \\
Quaoar's RA$^{(a)}$ &  \phantom{-}18 15 03.0717  $\pm$ 7 mas   \\
Quaoar's Dec$^{(a)}$ & -15 15 04.941\phantom{0}  $\pm$ 3 mas   \\

Quaoar's distance$^{(a)}$ &  41.840957 au   \\
Quaoar's Mag V & $\sim$18.9 \\
Shadow's velocity & -23.98 km/s \\
\hline
\hline 
\multicolumn{2}{l}{\rule{0pt}{3.0ex} $^{a}$ The used Quaoar ephemeris was the \textsc{nima\_v14} available}\\
\multicolumn{2}{l}{at the ERC Lucky Star Webpage.}\\

\end{tabular}\label{table:occinfo}
\end{center}
\end{table}

\begin{figure*}
\begin{center}
\includegraphics[width=0.75\textwidth]{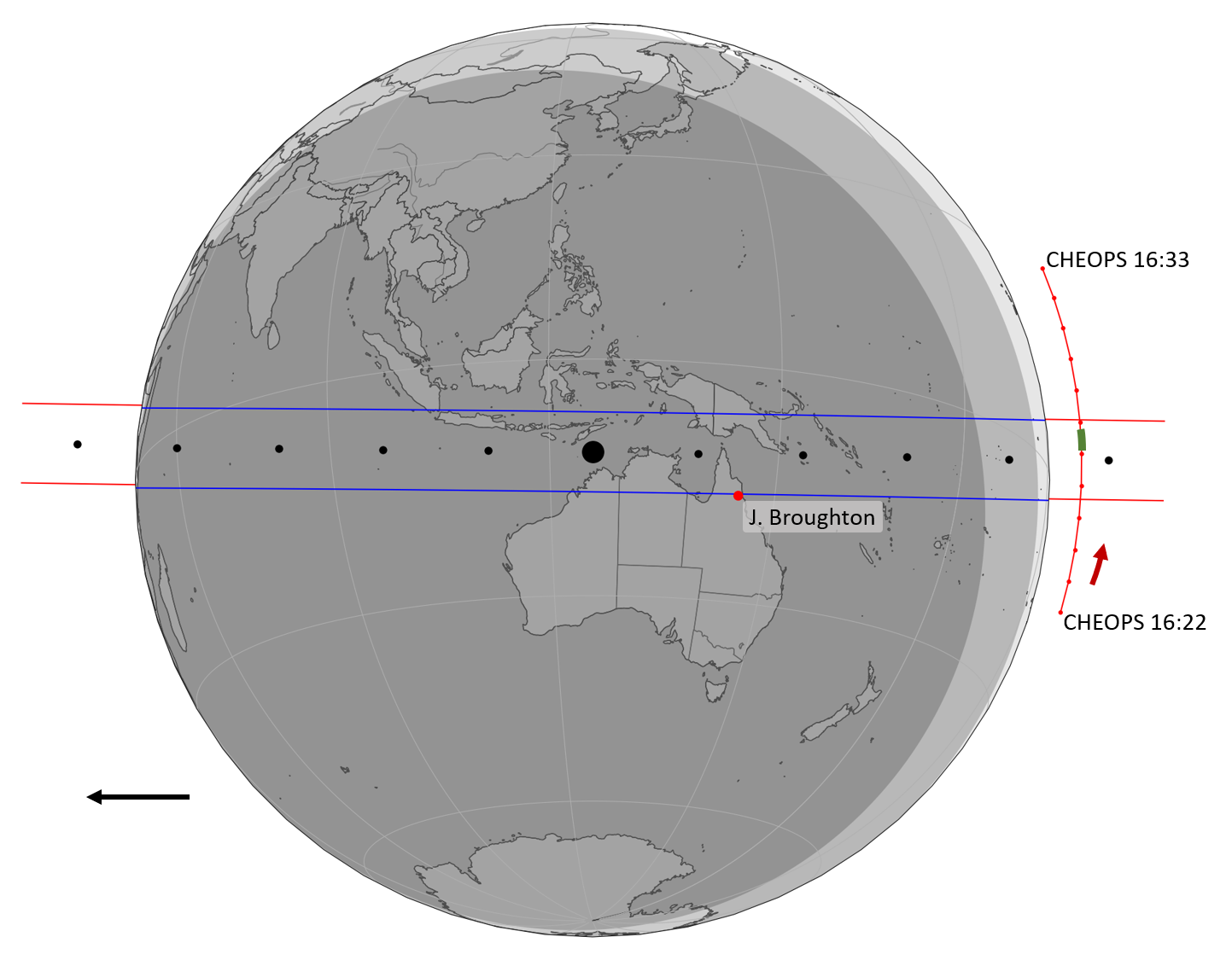}
\caption{Prediction map of the Quaoar event on 2020 June 11, the black arrow on the lower left corner shows the direction of the shadow’s movement. The blue lines stand for the shadow limits, and the black dots are the centre of the shadow, separated by one minute, with the biggest one representing the geocentric closest approach time. The red dot on the map represents the position of the ground-based observer who participated in this campaign. The red line on the right side of the map shows the projected position of CHEOPS from 16:22 to 16:33 UTC, two consecutive red dots are separated by one minute in time, and the red arrow emphasises the direction of motion. The green line shows when and where CHEOPS obtained the positive chord.}
\label{Fig:pred}
\end{center}
\end{figure*}

\Autoref{Fig:pred} shows the prediction map for the occultation event. The red line on the right side of the map shows the position of CHEOPS from 16:22 to 16:33 UTC, and the green segment shows when and where CHEOPS observed the occultation (positive chord). The red dot represents the position of the ground-based observer J. Broughton, in Mount Carbine, Australia (144$^\circ$ 52$'$ 29.4$''$ E, 16$^\circ$ 27$'$ 58.7$''$ S), who observed the star under a clear sky with a 25-cm telescope, but did not see the expected occultation, thus it is a negative chord.

The restituted CHEOPS ephemeris that was used contains its state vectors ($X$, $Y$, $Z$, $\dot{X}$, $\dot{Y}$, and $\dot{Z}$) relative to the geocentre, with axes in the ICRS, between 2020 June 11 06:51 UTC and 2020 June 12 23:56 UTC. The observation was done between 15:47:55.069 and 16:46:05.024 UTC with an exposure time of 3.0 seconds and a mean cycle of 3.024 seconds. It resulted in 1148 frames where the target star and the TNO were measured in the same aperture. CHEOPS ephemeris and the observed data are available in its archive\footnote{Webpage: \url{https://cheops-archive.astro.unige.ch/archive_browser/}}. 

\section{Light curve analysis and times}\label{sec:lightcurve}

CHEOPS was designed for exoplanet transit events, and the typical time resolution is 30 – 60s, and a 200-pixel diameter sub-image was downloaded for the photometry. However, to avoid saturation for brighter stars, shorter exposure times can be chosen, and the images were stacked on board before being downloaded. In addition, smaller images -- called imagettes -- with a size of only 50 pixels in diameter, are provided for each exposure. Because of the (deliberately)
de-focussed image, this is just enough to allow for photometry to enable high time resolution photometry. In the present case, the time resolution is 3s, and the reduction steps are the following: bias subtraction, flat-fielding, non-linearity correction, pointing jitter calculation, and photometry.

First, the bias was determined from the overscan data provided with each sub-image. The bias is stable, and no uncertainty was introduced by interpolation to the 14 times more frequent imagettes. Then, we applied flat-fielding, which depends on the wavelength due to CHEOPS's broad spectral response. Therefore, the applied flat field was weighted to match the spectral energy distribution of the star (in the present case, the effective temperature was 5040 K).

The needed non-linearity correction is a minor correction based on pre-launch measurements \citep{Deline_2019, Deline_2020, Futyan_2020}. For the background subtraction, we need to consider that the imagettes are too small to determine the background (basically the zodiacal light, but also the PSF wings of field stars). For this reason, sub-images were used for background calculation, and the background for each imagette was interpolated. The pointing jitter was evaluated by tracking the position of one of the three bright peaks in the PSF. The jitter amplitude was typically less than $\pm$ 0.5 pixels.

Finally, the photometry was done considering that the Field of View (FoV) around the target star is crowded. So, to minimise the contamination from field stars, we used an aperture mimicking the PSF of the star rather than using a standard circular aperture. The total number of pixels in this mask was 852.

Observations made on ground by J. Broughton's were analysed using the Platform for Reduction of Astronomical Images Automatically \citep[\texttt{PRAIA,}][]{Assafin_2011}. The occulted star was measured with a circular aperture and nearby stars were used as photometric calibrators to correct for sky fluctuations.

After normalising the light curve, we obtained the immersion and emersion times using Stellar Occultation Reduction and Analysis \citep[\textsc{sora}, ][]{SORA}\footnote{Webpage: \url{https://sora.readthedocs.io/}}. This software fits an occultation model that considers a sharp-edge box model convolved with Fresnel diffraction, the stellar diameter (projected at the body's distance), CCD bandwidth, and finite integration time. The light curves and the fitted models are available in CDS.

\begin{figure}
\begin{center}
\includegraphics[width=0.50\textwidth]{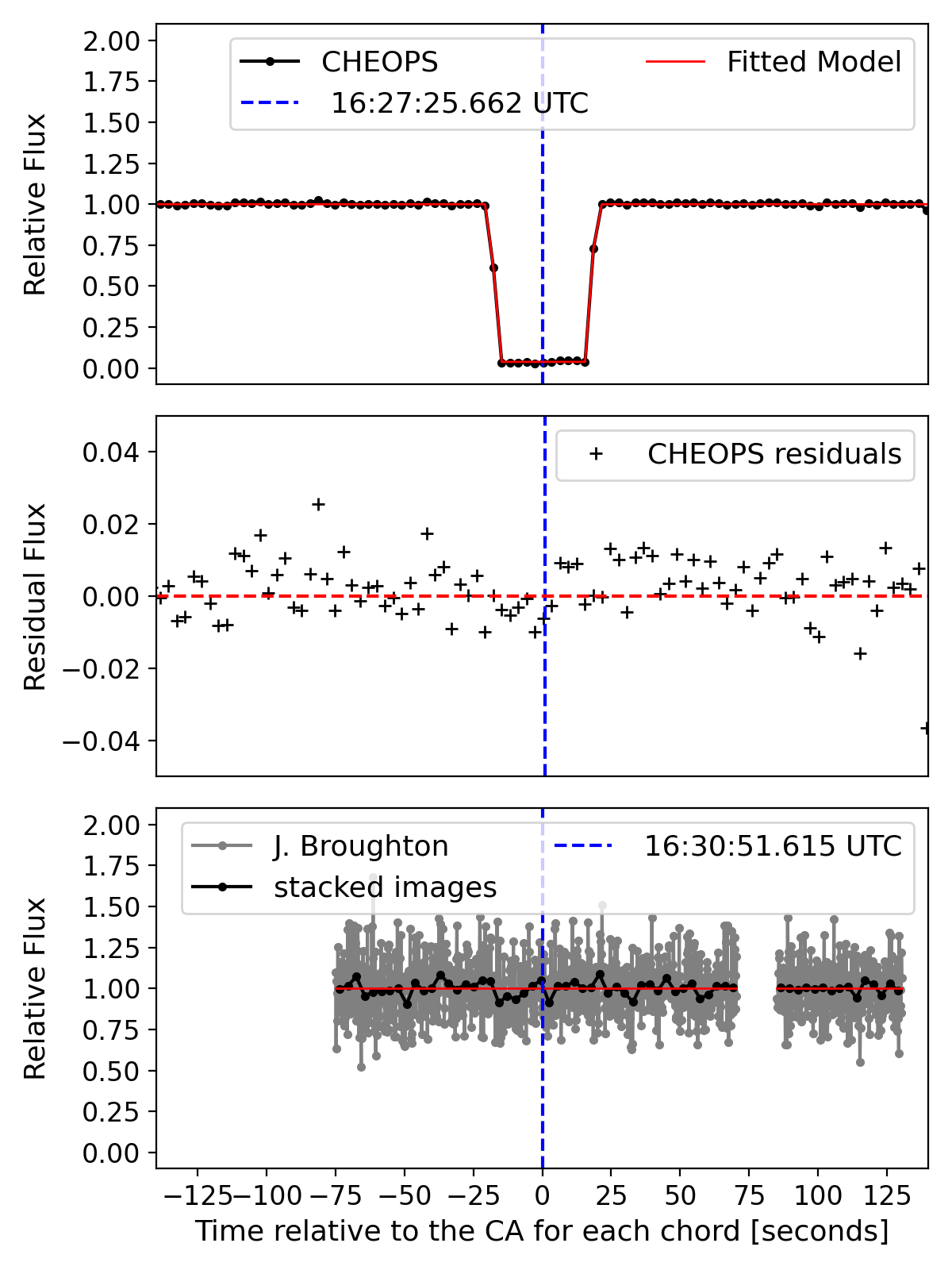}
\caption{Normalised light curves on 2020 June 11 during the Quaoar's stellar occultation relative to the closest approach (CA) time for each observer. The upper panel contains the CHEOPS light curve (black dots) and the fitted model (red line), and the middle panel contains CHEOPS residuals, in the sense of the observed flux minus the model. The bottom panel contains Broughton's negative light curve, where the grey line stands for the single images, and the black line stands for the stack of 19 images, resulting in a better signal-to-noise ratio.}
\label{Fig:lightcurve}
\end{center}
\end{figure}

\Autoref{Fig:lightcurve} contains the normalised light curves and the fitted models. The obtained immersion and emersion times were 16:27:08.199 $\pm$ 0.020 and 16:27:43.556 $\pm$ 0.020, resulting in a chord duration of 35.357 $\pm$ 0.040 seconds. The minimum $\chi^2$ per degree of freedom obtained was 0.945. CHEOPS light curve standard deviation outside the event was 0.76\%. Usually, only space telescopes can acquire such high photometric precision. Broughton's negative observation was separated into two blocks. The first was obtained between 16:29:36.64 and 16:32:02.08 UTC, and the second between 16:32:16.77 and 16:33:02.05 UTC, both with exposures of 0.16 seconds. Broughton's light curve has a standard deviation of 14.74\%. A fairer comparison can be made if we stack Broughton's light curve to have a similar exposure time as CHEOPS' light curve (e.g. 3 seconds); for that, we stacked 19 data points ($19~\times~0.16~s~=~3.04~s$) and achieved a standard deviation of 4.2\%.

The photometric accuracy obtained with CHEOPS for the occultation presented here makes this light curve an interesting case study in searching for an atmosphere around Quaoar. We determined an upper limit of a pure methane (CH$_4$) atmosphere using the ray-tracing method \citep{Dias-Oliveira_2015}, the obtained upper limit was 85 nbar. This large value is caused by the large exposure time (3 seconds) and the fact that all the disappearance and reappearance information happen during a single data point. Simulations where the star disappearance happens in the limit between two data points shows that we would have a detection limit of about 3 nbar in that scenario. The detailed analysis can be found in \Autoref{app:atm}.

\section{Geometric reconstruction of the event}\label{sec:Limb-fit}

The instants of disappearance and reappearance of the star behind the occulting body are the moments where the line-of-sight observer star matches the projected limb of the body. Therefore, the relative position between star and body at these moments gives the bi-dimensional position ($f$, $g$) of the limb relative to the body's centre in a plane perpendicular to the line of sight, the tangent plane. This projection was made with the \textsc{sora} package, and it results in a chord's size of 871.61 $\pm$ 0.99 km. 

The obtained chord is compatible with the area equivalent diameter obtained by stellar occultations \cite[1110 km,][]{Braga-Ribas_2013} and with the diameter derived from thermal data \citep[1074 km,][]{Fornasier_2013}. Each chord extremity is a point at which we can fit the apparent shape of the occulting object. As only one chord was recorded, we fitted the centre of the figure ($f_0$, $g_0$), allowing the shape to vary within the Maclaurin spheroid solution proposed by \cite{Braga-Ribas_2013}, with an equatorial radius of $581^{+12}_{-8}$ km, apparent oblateness varying between 0 and 0.114, and a pole position angle between 0 and 180 degrees. This fit was done using \textsc{sora}, with a method that is fully described in \cite{SORA}. The negative chord was used to eliminate solutions that would result in a positive detection of the occultation by J. Broughton.

\begin{figure}[h]
\begin{center}
\includegraphics[width=0.47\textwidth]{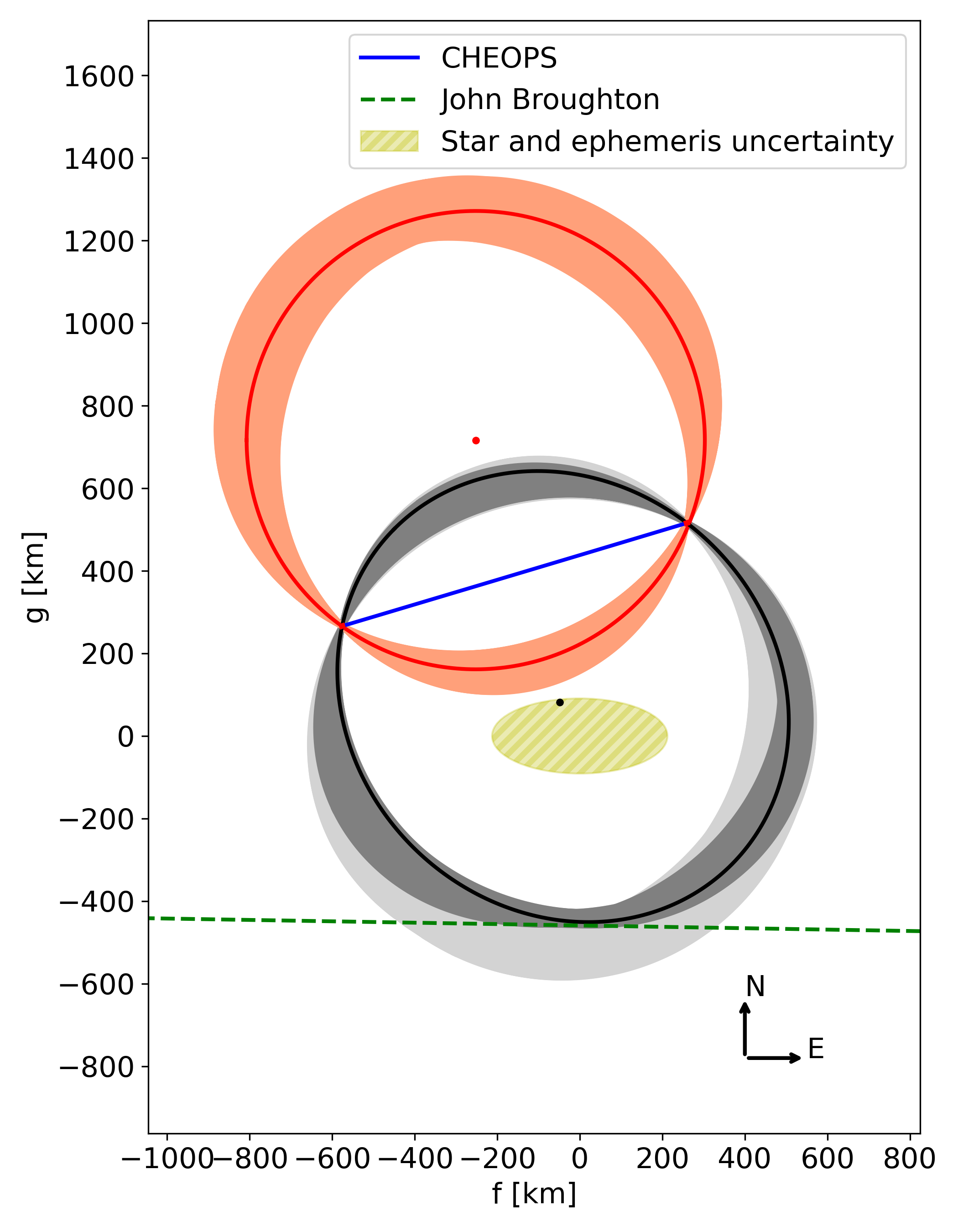}
\caption{Best fitted limb to the CHEOPS chord (in blue) and ellipses within their respective 1-$\sigma$ region. Two solutions can be obtained, one with the centre north of the chord (in red), and another to the south (in black). The black line stands for the preferred resulted central position of Quaoar ($f_0,~g_0$) considering the centre to the south of the chord. The red line is the unlikely northern solution. The solution in black is the preferred solution as it agrees more with the expected orbit of Quaoar based on the \textsc{nima} ephemeris and its uncertainty (yellow dashed region). The green dashed line corresponds to the negative observation used to eliminate some of the solutions that cross it (light grey ellipses).}
\label{Fig:ellipse_fit}
\end{center}
\end{figure}

\Autoref{Fig:ellipse_fit} shows the best-fitted ellipses and their respective 1$\sigma$ region as representatives of Quaoar's limb. Two solutions can be obtained, one with the centre north of the chord (at $f_0$~=~$-$252 km and $g_0$~=~ $-$716 km, in red), and the second with the centre to the south (at $f_0$~=~$-$48 km and $g_0$~=~ $+$81 km, in black). As can be seen in \Autoref{table:occinfo}, the \textsc{nima} software obtained an uncertainty of $\sim$212 km in RA and $\sim$90 km in December for Quaoar ephemeris. Moreover, it is unlikely that the north solution is the correct one as it is not consistent with the knowledge we have about Quaoar’s orbit. Finally, we can shift the reference frame to the geocentre and determine Quaoar's astrometric geocentric position on 2020\ June 11 at 16:27:25.500 UTC as

\begin{eqnarray}
RA &=&  \phantom{-}18^h~15'~03''.0715863 \pm 0.864~\textrm{mas}, \nonumber \\
DEC &=& -15^{\circ}~15'~04''.937760~\pm~0.729~\textrm{mas}. \nonumber
\end{eqnarray}
This position can be used to improve Quaoar's ephemeris, which in turn helps us to predict future, accurate stellar occultations for this object.

As this was a single chord event, no further constraint on Quaoar's shape was obtained. However, the resulting pole position angle for Quaoar was restrained to values between 14.2 $\pm$ 39.8 degrees due to J. Broughton's negative chord. This orientation is in agreement with Weywot's orbital pole position for the occultation date of $-$5.2 degrees, using the pole orientation published by \cite{Vachier_2012}.

\section{Discussion}\label{sec:discussion}

Stellar occultations are transient events that allow us to obtain physical parameters of the bodies in our Solar System. Networks of ground-based telescopes have been used to observe such events and derive many relevant results. Stellar occultations have also been observed by spacecrafts, such as the Cassini mission \citep{Li_2014}. However, these were mostly in loco observations with events involving the objects the spacecrafts were visiting. 

From space telescopes orbiting Earth, HST observed a stellar occultation by Saturn and its rings in 1991, as shown in \cite{Elliot1993}. Furthermore, analysing HST Fine Guidance Sensors, two serendipitous occultations by small Oort cloud objects were reported by \cite{Schlichting2009} and \cite{Schlichting2012}.

For the first time, we report on the detection of a stellar occultation by a minor body as the primary target of the observation of a space telescope orbiting Earth. We obtained a non-diametrical chord of 871.61 $\pm$ 0.99 km for the TNO Quaoar, consistent with its previously published size. As a result, we detected limits of a global methane atmosphere around Quaoar. Finally, we obtained an astrometric position with an uncertainty better than 1~mas.

It is important to highlight that more ground-based stations could observe this occultation. Combining space- and ground-based observation can provide more constraints for the limb fitting and improve the result. Also, ground-based observations will detect chords that are usually parallel to each other, while the orientation of a space telescope will differ, which can be useful in the limb fitting as shown in \autoref{Fig:ellipse_fit}. Finally, functions in the \textsc{sora} package can be used to prepare automatic prediction pipelines based on the ephemeris of the space telescope and occulting objects, as presented in \Autoref{app:future_events}. 

The space observation of the Quaoar occultation presented in this paper is the first of its kind, and it serves as a proof of concept for future campaigns. Space telescopes and CubeSat telescopes can be used in conjunction with ground-based stations to detect stellar occultations. Space telescopes are not affected by weather and they extend the range of the Earth where an occultation can be observed. They can also help overcome the circumstances where the shadow would mostly cross the ocean and limited ground-based observations can be made, thus increasing the number of events observed. It is important to highlight that CHEOPS's diameter is modest, but it allows for a photometric accuracy equivalent to what ground-based telescopes with much larger apertures would obtain.

Moreover, as discussed by \cite{Santos-Sanz_2016}, the prospect for the James Webb Space Telescope (JWST\footnote{Webpage: \url{https://jwst.nasa.gov/content/webbLaunch/index.html}}) is impressive. JWST's large primary mirror (6.5 metres in diameter) with a temporal sampling of 20 frames per second will allow for the precise characterisation of material around the small bodies in our Solar System using stellar occultations, which would include detecting faint rings and thin atmospheres within the occulting bodies.

\begin{acknowledgements}
This work was carried out within the “Lucky Star" umbrella that agglomerates the efforts of the Paris, Granada and Rio teams, which is funded by the European Research Council under the European Community’s H2020 (ERC Grant Agreement No. 669416).
CHEOPS is an ESA mission in partnership with Switzerland with important contributions to the payload and the ground segment from Austria, Belgium, France, Germany, Hungary, Italy, Portugal, Spain, Sweden, and the United Kingdom. The CHEOPS Consortium would like to gratefully acknowledge the support received by all the agencies, offices, universities, and industries involved. Their flexibility and willingness to explore new approaches were essential to the success of this mission.
This research made use of \textsc{sora}, a python package for stellar occultations reduction and analysis, developed with the support of ERC Lucky Star and LIneA/Brazil, within the collaboration of Rio-Paris-Granada teams.
This work has made use of data from the European Space Agency (ESA) mission Gaia (\url{https://www.cosmos.esa.int/gaia}), processed by the Gaia Data Processing and Analysis Consortium (DPAC, https://www.cosmos.esa.int/web/gaia/dpac/consortium).
This study was financed in part by the National Institute of Science and Technology of the e-Universe project (INCT do e-Universo, CNPq grant 465376/2014-2).
The following authors acknowledge the respective 
i) CNPq grants: 
BEM 150612/2020-6; 
FB-R 314772/2020-0; 
ii) CAPES/Cofecub grant:
BEM 394/2016-05.
ii) FAPESP grants: 
ARGJr 2018/11239-8; 
GBr, VNa, IPa, GPi, RRa, and GSc acknowledge support from CHEOPS ASI-INAF agreement n. 2019-29-HH.0. 
ABr was supported by the SNSA.
ACC acknowledges support from STFC consolidated grant numbers ST/R000824/1 and ST/V000861/1, and UKSA grant number ST/R003203/1.
KGI is the ESA CHEOPS Project Scientist and is responsible for the ESA CHEOPS Guest Observers Programme. She does not participate in, or contribute to, the definition of the Guaranteed Time Programme of the CHEOPS mission through which observations described in this paper have been taken, nor to any aspect of target selection for the programme.
ACC and TW acknowledge support from STFC consolidated grant numbers ST/R000824/1 and ST/V000861/1, and UKSA grant number ST/R003203/1.
YA and MJH acknowledge the support of the Swiss National Fund under grant 200020\_172746.
We acknowledge support from the Spanish Ministry of Science and Innovation and the European Regional Development Fund through grants ESP2016-80435-C2-1-R, ESP2016-80435-C2-2-R, PGC2018-098153-B-C33, PGC2018-098153-B-C31, ESP2017-87676-C5-1-R, MDM-2017-0737 Unidad de Excelencia Maria de Maeztu-Centro de Astrobiologí­a (INTA-CSIC), as well as the support of the Generalitat de Catalunya/CERCA programme. The MOC activities have been supported by the ESA contract No. 4000124370.
S.C.C.B. acknowledges support from FCT through FCT contracts nr. IF/01312/2014/CP1215/CT0004.
XB, SC, DG, MF and JL acknowledge their role as ESA-appointed CHEOPS science team members.
This project was supported by the CNES.
The Belgian participation to CHEOPS has been supported by the Belgian Federal Science Policy Office (BELSPO) in the framework of the PRODEX Program, and by the University of Liège through an ARC grant for Concerted Research Actions financed by the Wallonia-Brussels Federation.
L.D. is an F.R.S.-FNRS Postdoctoral Researcher.
This work was supported by FCT - Fundação para a Ciência e a Tecnologia through national funds and by FEDER through COMPETE2020 - Programa Operacional Competitividade e Internacionalizacão by these grants: UID/FIS/04434/2019, UIDB/04434/2020, UIDP/04434/2020, PTDC/FIS-AST/32113/2017 \& POCI-01-0145-FEDER- 032113, PTDC/FIS-AST/28953/2017 \& POCI-01-0145-FEDER-028953, PTDC/FIS-AST/28987/2017 \& POCI-01-0145-FEDER-028987, O.D.S.D. is supported in the form of work contract (DL 57/2016/CP1364/CT0004) funded by national funds through FCT.
P.S-S. acknowledges financial support by the Spanish grant AYA-RTI2018-098657-J-I00 “LEO-SBNAF” (MCIU/AEI/FEDER, UE). P.S-S. and J.L.O. acknowledge financial support from the State Agency for Research of the Spanish MCIU through the “Center of Excellence Severo Ochoa” award for the Instituto de Astrofísica de Andalucía (SEV-2017-0709), they also acknowledge the financial support by the Spanish grants AYA-2017-84637-R and PID2020-112789GB-I00, and the Proyectos de Excelencia de la Junta de Andalucía 2012-FQM1776 and PY20-01309.
B.-O.D. acknowledges support from the Swiss National Science Foundation (PP00P2-190080).
This project has received funding from the European Research Council (ERC) under the European Union’s Horizon 2020 research and innovation programme (project {\sc Four Aces}.
grant agreement No 724427). It has also been carried out in the frame of the National Centre for Competence in Research PlanetS supported by the Swiss National Science Foundation (SNSF). DE acknowledges financial support from the Swiss National Science Foundation for project 200021\_200726.
MF and CMP gratefully acknowledge the support of the Swedish National Space Agency (DNR 65/19, 174/18).
DG gratefully acknowledges financial support from the CRT foundation under Grant No. 2018.2323 ``Gaseous or rocky? Unveiling the nature of small worlds''.
M.G. is an F.R.S.-FNRS Senior Research Associate.
SH gratefully acknowledges CNES funding through the grant 837319.
This work was granted access to the HPC resources of MesoPSL financed by the Region Ile de France and the project Equip@Meso (reference ANR-10-EQPX-29-01) of the programme Investissements d'Avenir supervised by the Agence Nationale pour la Recherche.
ML acknowledges support of the Swiss National Science Foundation under grant number PCEFP2\_194576.
PM acknowledges support from STFC research grant number ST/M001040/1.
This work was also partially supported by a grant from the Simons Foundation (PI Queloz, grant number 327127).
IRI acknowledges support from the Spanish Ministry of Science and Innovation and the European Regional Development Fund through grant PGC2018-098153-B- C33, as well as the support of the Generalitat de Catalunya/CERCA programme.
S.G.S. acknowledge support from FCT through FCT contract nr. CEECIND/00826/2018 and POPH/FSE (EC).
GyMSz acknowledges the support of the Hungarian National Research, Development and Innovation Office (NKFIH) grant K-125015, a a PRODEX Experiment Agreement No. 4000137122, the Lend\"ulet LP2018-7/2021 grant of the Hungarian Academy of Science and the support of the city of Szombathely.
VVG. is an F.R.S-FNRS Research Associate.
NAW acknowledges UKSA grant ST/R004838/1.

\end{acknowledgements}

\bibliographystyle{aa} 
\bibliography{ref} 

\appendix

\section{Quaoar's atmospheric limit based on CHEOPS data} \label{app:atm}

\begin{figure}[h]
    \centering
    \includegraphics[width=1.0\linewidth]{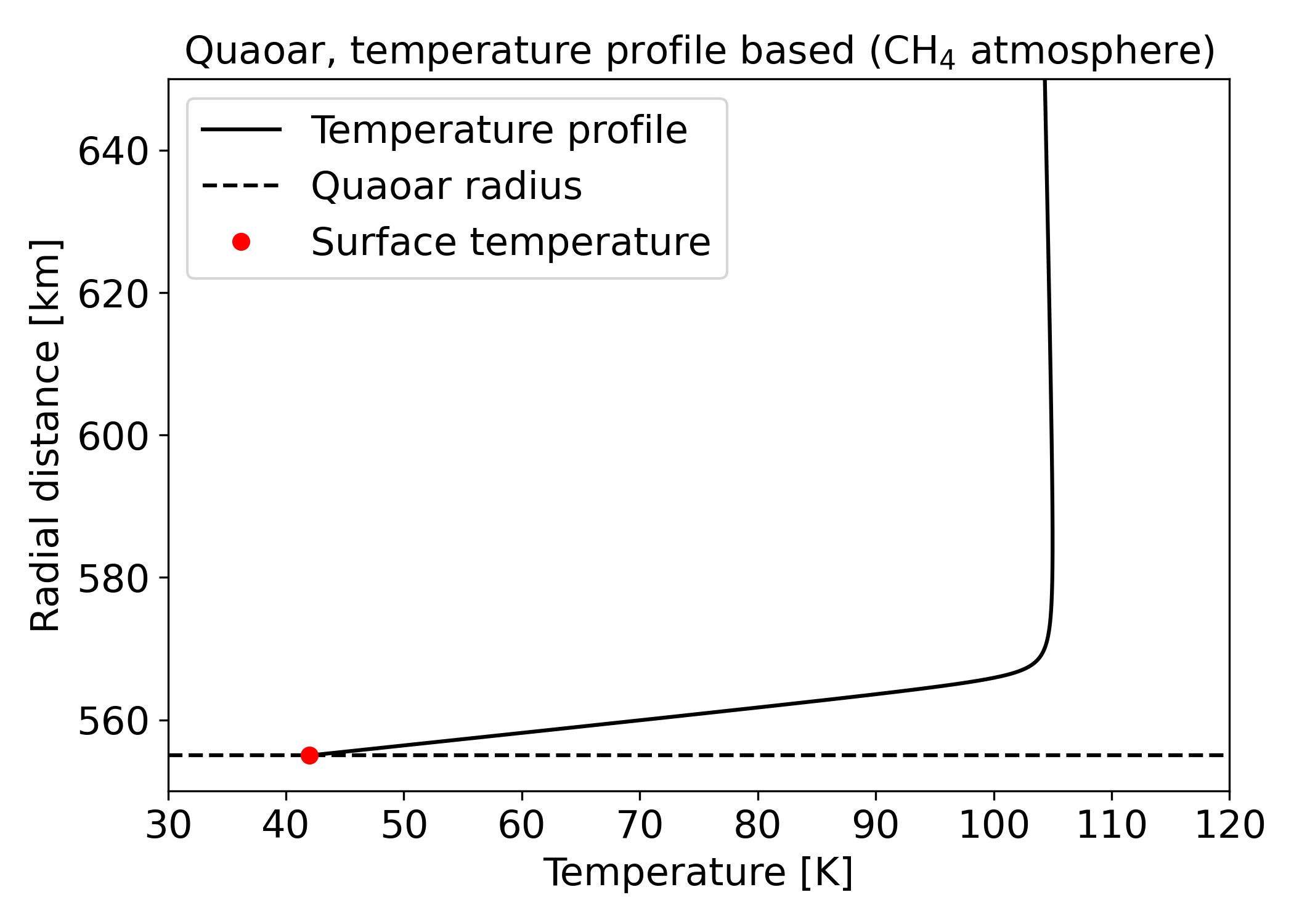}
    \caption{Assumed temperature profile in Kelvin over radial distance in kilometres of a pure methane (CH$_4$) atmosphere.}
    \label{fig:atm_T_r}
\end{figure}

Using CHEOPS data for the 11 June 2020 occultation, we determined an upper limit of Quaoar's atmosphere. The model assumed was a pure methane (CH$_4$) atmosphere with a temperature gradient at the surface of $(dT/dz)_{\sf surface}~=~5.7$~K~km$^{-1}$, and an isothermal upper branch at 102~K which is reached in about 10~km, see \Autoref{fig:atm_T_r}. The tested model is consistent with the ones used in \cite{Braga-Ribas_2013} and \cite{Arimatsu_2019}. Quaoar was considered to be a sphere with a radius of 555 km; any departure from that should be negligible at this level of estimation.

A ray-tracing method (See \cite{Dias-Oliveira_2015} and references therein) was used to compute synthetic light curves convoluted with the instrumental response (exposure time of 3 seconds). We varied the surface pressure ($P_{surface}$, in nbar) and the closest approach (CA, in kilometres) between the star's and Quaoar's central position, and we performed Chi-squared ($\chi^2$) statistical tests to check the detection limit of a potential methane atmosphere.

\begin{figure}[h]
    \centering
    \includegraphics[width=1.0\linewidth]{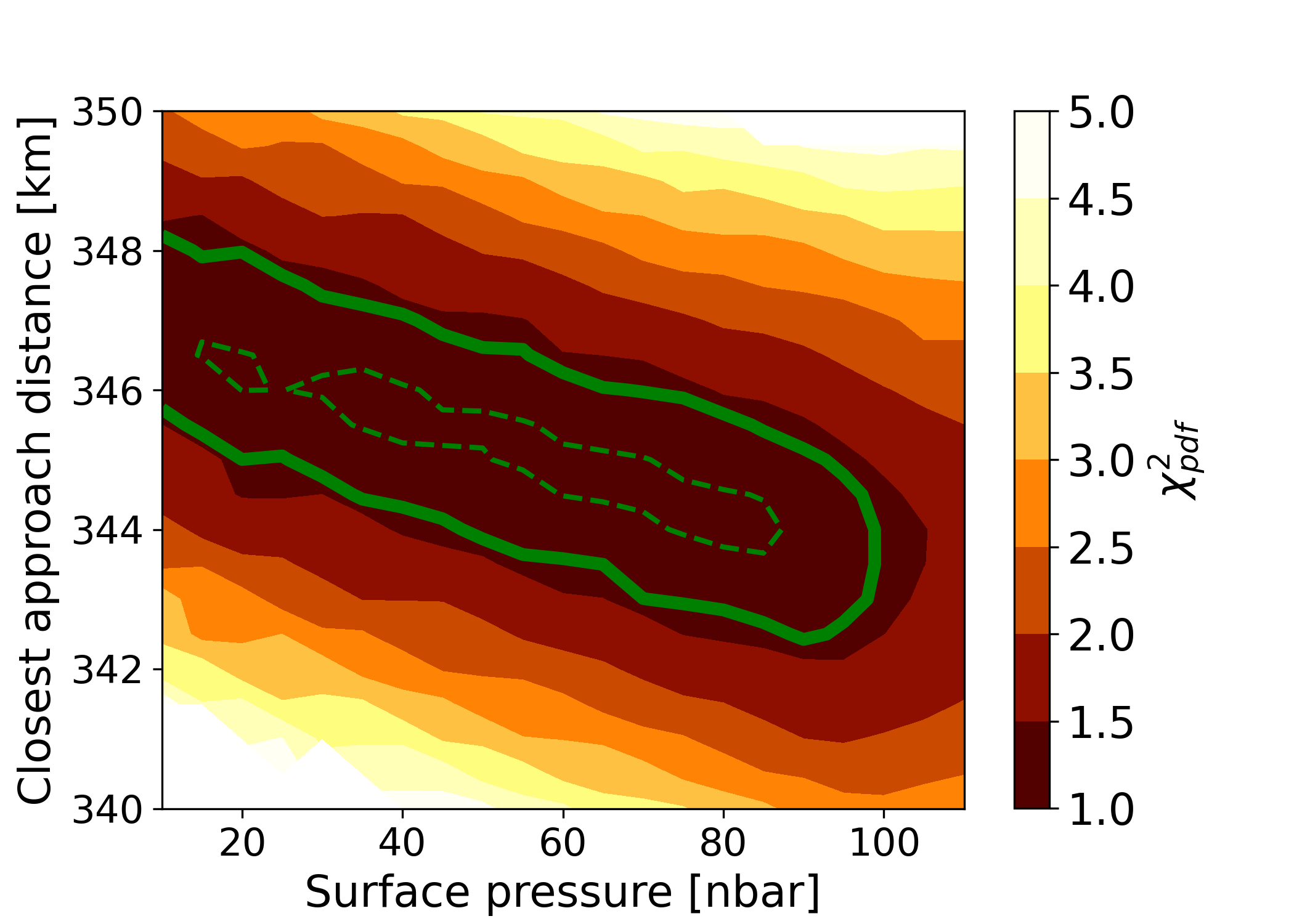}
    \caption{Chi-squared per degree of freedom ($\chi^2_{pdf}$) map vs. surface pressure and CA distance. The inner dashed (resp. outer solid) green level curve is the 1$\sigma$ (resp. 3$\sigma$) domain.}
    \label{fig:atm_chi2_map}
\end{figure}

This analysis resulted in a surface pressure between 15 and 85 nbars for the 1$\sigma$ confidence level, and values between 0 and 95 nbars for 3$\sigma$. So no significant value was obtained and only the 85 nbar as an upper limit should be considered. Nonetheless, this limit is larger than the upper limits determined by \cite{Braga-Ribas_2013} of 20 nbar, and by \cite{Arimatsu_2019} of 6 nbar. This is caused by the large exposure time (3 seconds, about 75 km in radial distance in the sky plane) and the fact that all the disappearance and reappearance information occurred within one exposure, so the various models (with different pressure) are all contained in that exposure. That makes the discrimination between pressures ineffective.

As a test, we simulated data where the star disappearance and reappearance  happens just at the limit between two exposures (the best case). This simulation resulted in an upper limit of only 3 nbar, assuming the same model. Is important to highlight that a smaller spatial resolution (e.g. smaller temporal resolution, smaller event velocity, or grazing occultations) would imply an even better result due to the photometric accuracy.

\section{Prediction of stellar occultations for CHEOPS} \label{app:future_events}

\begin{figure}[h]
    \centering
    \includegraphics[width=1.0\linewidth]{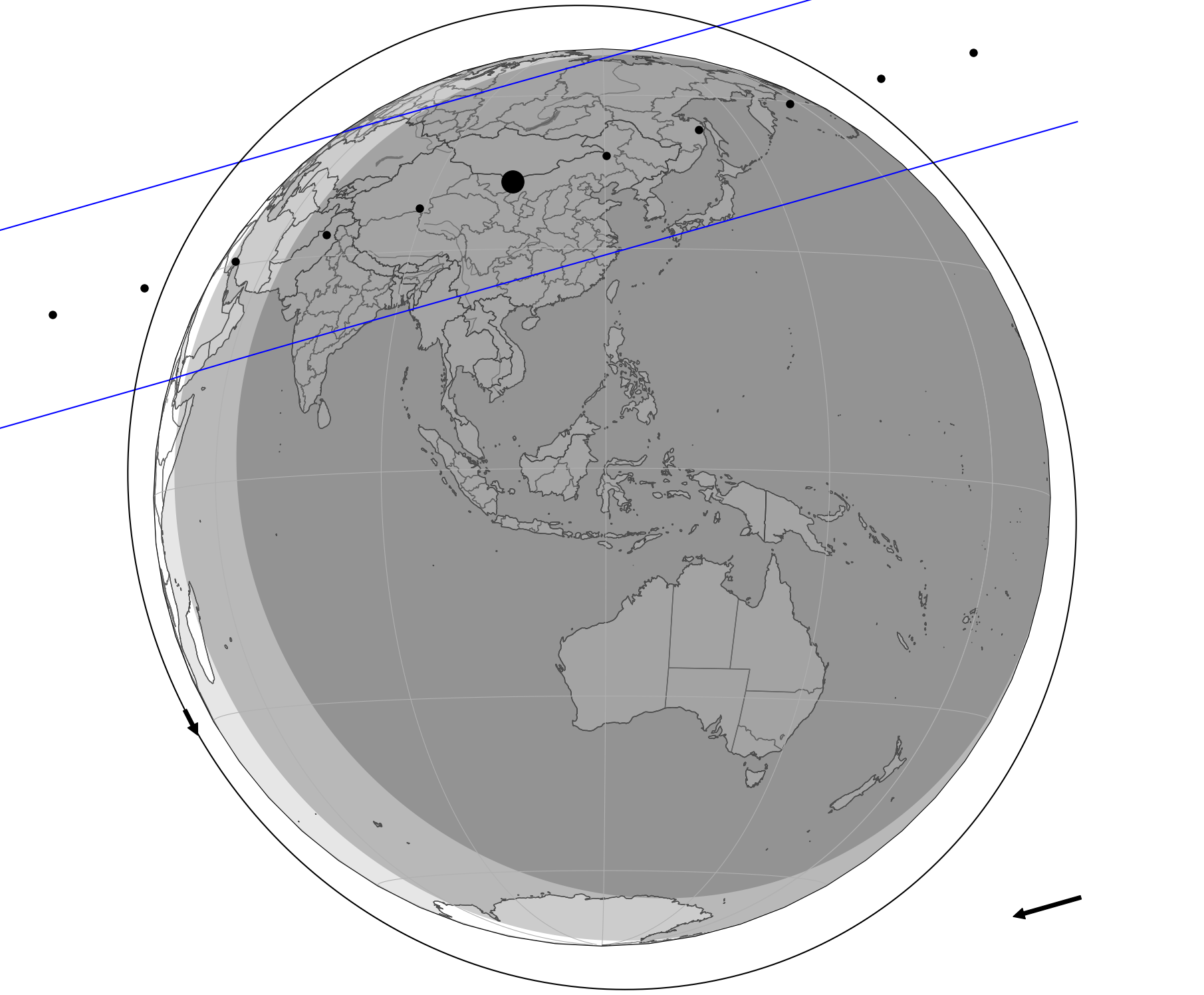}
    \caption{Prediction map of a stellar occultation by Triton on 2022 October 06. The path of CHEOPS is shown as the circle just above the Earth, with an arrow indicating its sense of motion.}
    \label{fig:occ_triton}
\end{figure}

\begin{figure}
    \centering
    \includegraphics[width=1.0\linewidth]{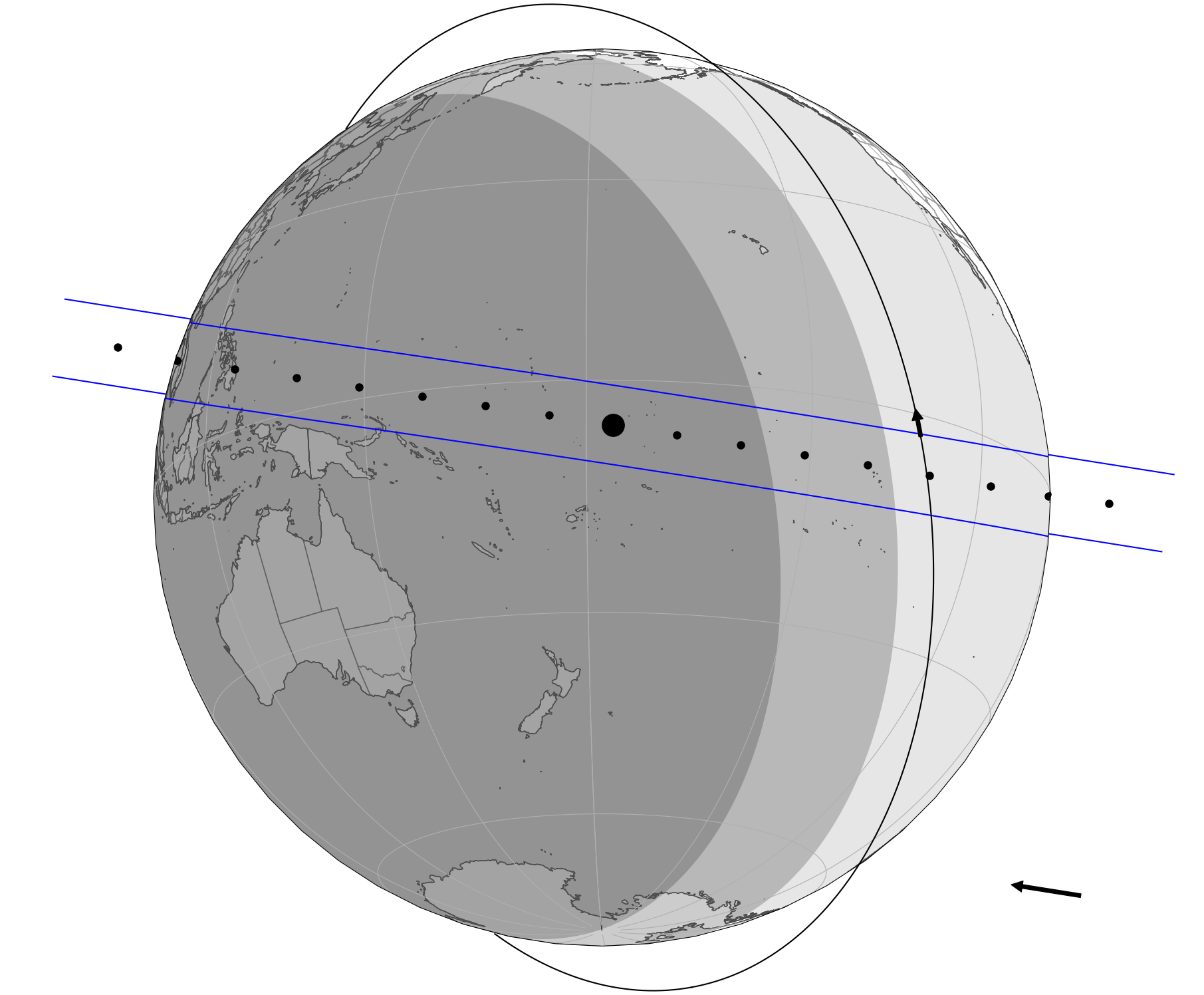}
    \caption{Similar to \autoref{fig:occ_triton}, but for the occultation by Quaoar on 2023 May 10.}
    \label{fig:occ_quaoar1}
\end{figure}

\begin{figure}
    \centering
    \includegraphics[width=1.0\linewidth]{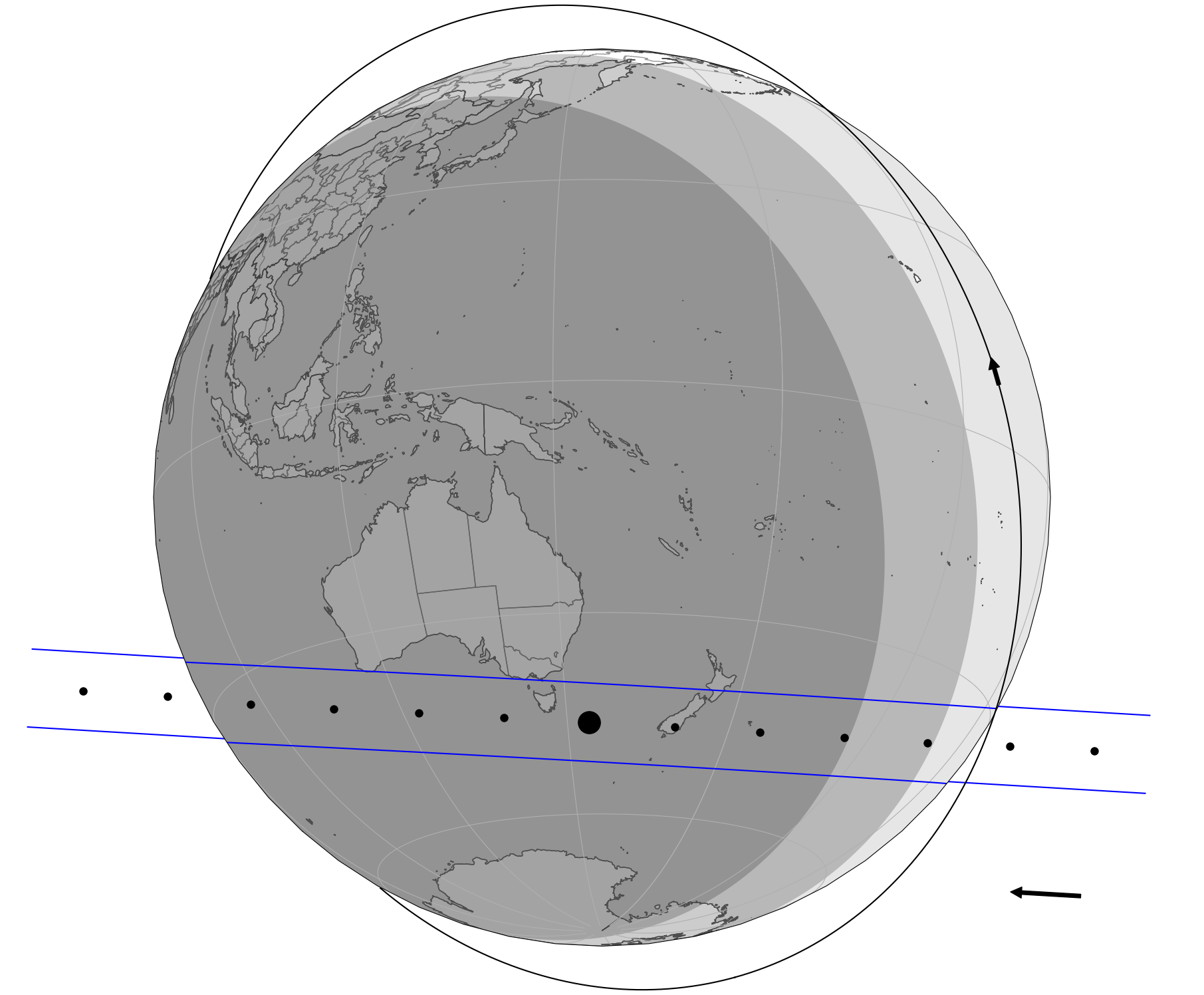}
    \caption{Similar to \autoref{fig:occ_triton}, but for the occultation by Quaoar on 2023 May 26.}
    \label{fig:occ_quaoar2}
\end{figure}

\begin{figure}
    \centering
    \includegraphics[width=1.0\linewidth]{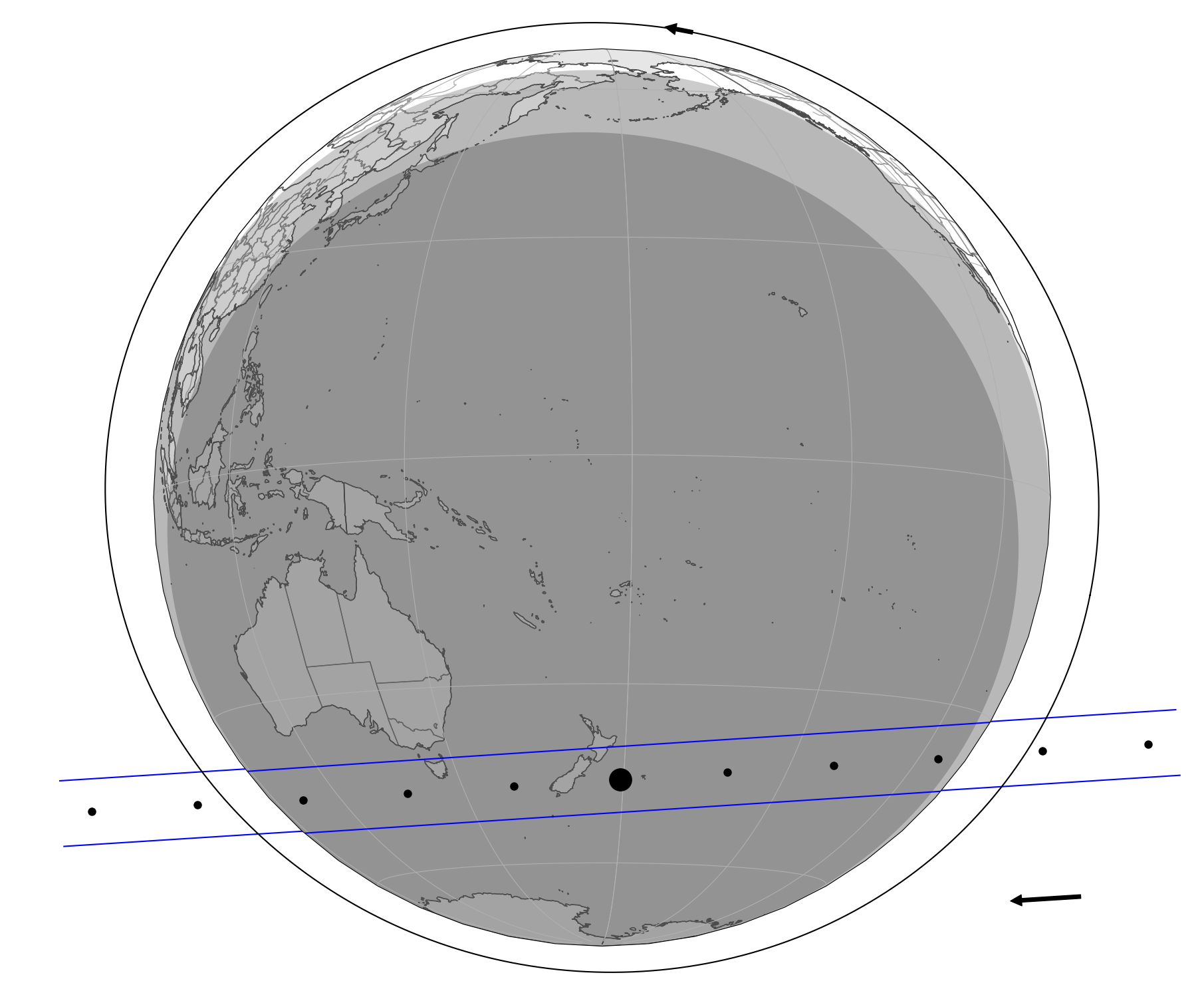}
    \caption{Similar to \autoref{fig:occ_triton}, but for the occultation by 2002 MS4 on 2023 July 03.}
    \label{fig:occ_2002ms4}
\end{figure}

\begin{figure}
    \centering
    \includegraphics[width=1.0\linewidth]{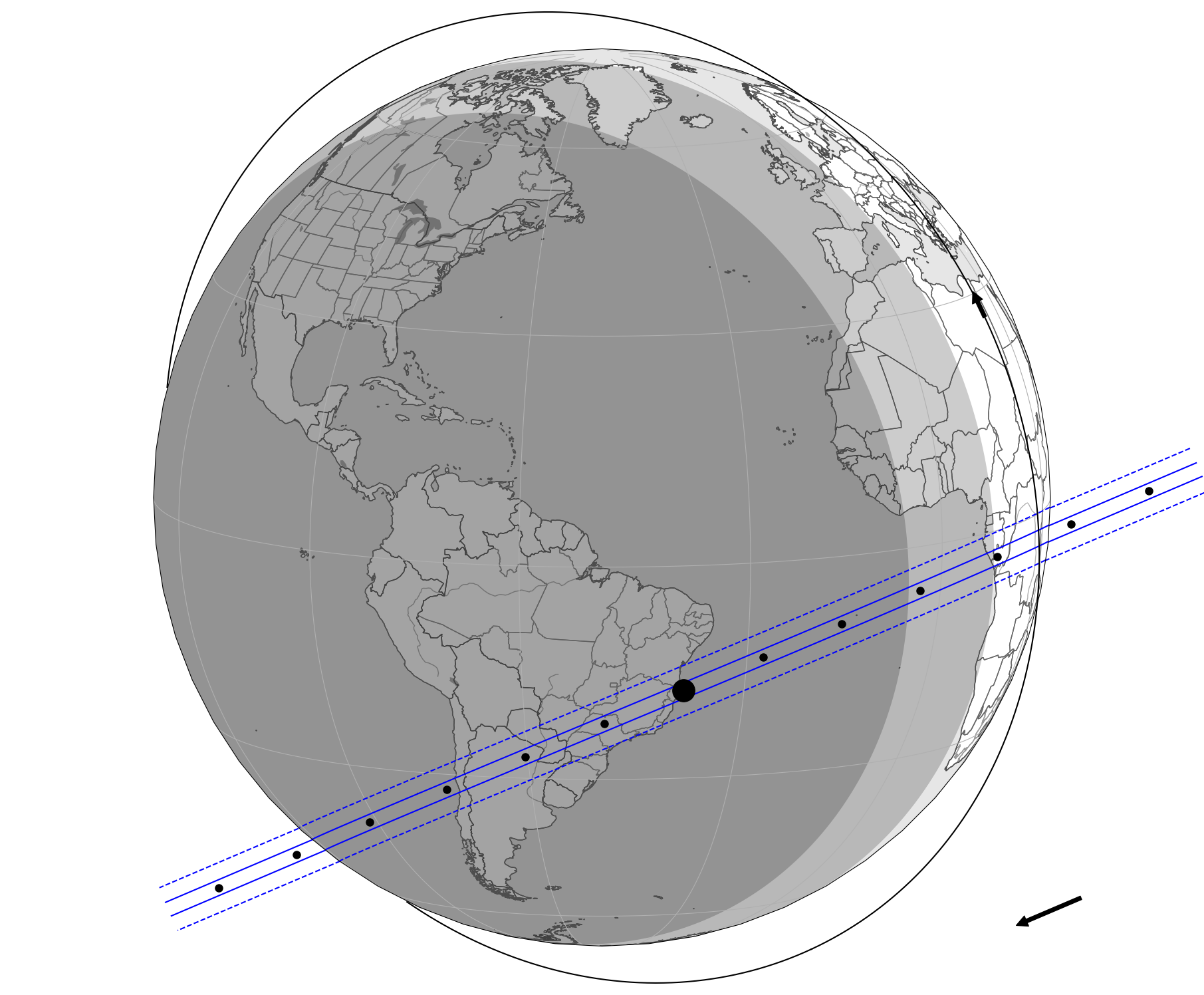}
    \caption{Similar to \autoref{fig:occ_triton}, but for the occultation by Quaoar on 2023 September 10.}
    \label{fig:occ_chiron}
\end{figure}

With the success achieved in observing the stellar occultation by Quaoar, we predicted events to be observed by CHEOPS in the next months. We were granted time to observe five stellar occultations by objects in the outer Solar System. That was done in the context of the third Announcement of Opportunity. The occultation prediction maps are shown in \Autoref{fig:occ_triton}-\ref{fig:occ_chiron}. The path of CHEOPS on space is given as the circle around Earth. Given the uncertainty as to the exact position of CHEOPS at the occultation epoch, the predictions were made by computing the probability of CHEOPS to observe the event, considering its orbit is kept in a polar orbit with a movement just above the terminator of the Earth, and fixing its orbital velocity. Ground-based campaigns are being organised, and observations are planned to be made from Earth to complement CHEOPS' observation.

\end{document}